\documentclass[aps,twocolumn,superscriptaddress,longbibliography,nofootinbib,floatfix,showkeys]{revtex4-2}
\usepackage[utf8]{inputenc}
\usepackage[T1]{fontenc}

\usepackage{amsmath}
\usepackage{amssymb}
\usepackage{graphicx}
\usepackage{xcolor}
\usepackage{subfigure}

\usepackage[italicdiff]{physics}
\usepackage[protrusion=true,expansion=true]{microtype}
\usepackage[colorlinks=True,allcolors=blue]{hyperref}
\usepackage{orcidlink}
\usepackage{cleveref}

\allowdisplaybreaks

\newcommand{\nextstep}{\nonumber\\[0.5ex]}
\newcommand{\nextline}{\\[0.5ex]}
\newcommand{\expect}[1]{\langle #1 \rangle}
\newcommand{\h}{\hat} 
\newcommand{\ha}{\hat{a}}
\newcommand{\da}{\downarrow}
\newcommand{\ua}{\uparrow}
\newcommand{\bvec}[1]{\mathbf{#1}}

\newcommand{\UGxICTQT}{International Centre for Theory of Quantum Technologies, University of Gda\'nsk, 80-309 Gda\'nsk, Poland}

\newcommand{\UGxIFTiA}{Institute of Theoretical Physics and Astrophysics, Faculty of Mathematics, Physics and Informatics, University of Gda\'nsk, 80-308 Gda\'nsk, Poland}

\newcommand{\UCRxPhysics}{Laboratorio de F\'isica Te\'orica y Computacional, Escuela de F\'isica, Universidad de Costa Rica, 11501-2060 San Jos\'e, Costa Rica}

\begin{document}

\title{Open quantum dynamics of Josephson charge pumps}

\author{Ankit Kumar\,\orcidlink{0000-0003-3639-6468}}
\email{kumar.ankit.vyas@gmail.com}
\affiliation{\UGxICTQT}
\affiliation{\UGxIFTiA}

\author{Luis Cort\,\orcidlink{0000-0001-5581-4333}}
\email{luis.cort-barrada@ug.edu.pl}
\affiliation{\UGxICTQT}

\author{Marcin {\L}obejko\,\orcidlink{0000-0002-7159-5502}}
\email{marcin.lobejko@ug.edu.pl}
\affiliation{\UGxIFTiA}

\author{Alejandro Jenkins\,\orcidlink{0000-0002-4463-4633}}
\email{alejandro.jenkins@ucr.ac.cr}
\affiliation{\UGxICTQT}
\affiliation{\UCRxPhysics}

\author{Micha{\l} Horodecki\,\orcidlink{0000-0002-0446-3059}}
\email{michal.horodecki@ug.edu.pl}
\affiliation{\UGxICTQT}
\affiliation{\UGxIFTiA}

\begin{abstract}
We investigate the macroscopic dynamics of Josephson charge pumps in the light of Alicki et al.'s theoretical description of the Josephson junction as an open quantum system described by a Markovian master equation. Once the electrostatic interaction between the terminals is taken into account via nonlinear capacitive terms in the Hamiltonian, we find that the resulting description of pumping is physically reasonable and in good qualitative agreement with experimental observations.  We comment on how this approach relates to other theoretical treatments of quantum pumps based on time-dependent potentials or scattering amplitudes. We also highlight the significance of our results in the broader context of the dynamics of charge pumping by active systems.
\end{abstract}

\maketitle

\section{Introduction}
\label{sec:intro}

A nanoscale pump is a miniature machine that utilizes quantum mechanical operations in nanostructures to enable a controlled and quantized charge transport~\cite{Thouless1983,Buttiker1994,Brouwer1998,Zhou1999}. 
As its name suggests, such a device is capable of moving only a few electrons at a time, thereby pumping only a minuscule amount of current per cycle of operation~\cite{Pothier_1992,Martinis_1994,Fletcher_2003,Ebbecke_2004,Fuhrer_2007,Pekola2008,Buitelaar_2008,Kaestner_2008,Giazotto2011}. 
One of the categories of such devices is the superconducting charge pump, in which Cooper pairs (rather than single electrons) are transported across a Josephson junction (JJ), without any external bias applied to it~\cite{Josephson1962}. 

The first experimental implementation of such a superconducting charge pump was reported by Geerligs et al.~\cite{Geerligs_1991}. The setup consisted of a linear chain of three JJs and resulted in a picoamp current. A very similar setup was later studied by Toppari et al.~\cite{Toppari_2004}, where they observed currents of the same order of magnitude. 
The output current was increased by designing the so-called ``sluice'', which resulted in an accurate pumping up to the nanoampere scale~\cite{Niskanen_2003,Niskanen_2005,Vartiainen_2007}.
The setup realized in \cite{Giazotto2011} was conceptually different from the standard scenarios: the pumping was not executed by means of periodic change of external parameters, but with a time dependence of the superconducting phase. The authors reported a picoampere current through an unbiased nanowire embedded in a superconducting quantum interference device (SQUID).
Due to an increasing ability to manipulate charges at the nanoscale, quantum pumps find numerous applications in the development of precision quantum standards~\cite{Blumenthal_2007, Kemppinen_2009, Giblin_2012, Camarota_2012}. 
It is therefore necessary to improve our understanding of the dynamics of such devices that will underpin future quantum technologies~\cite{Burrello_Thouless}.

In this work we study such pump-type devices by adapting the theoretical treatment of the JJ as an open quantum system that was recently proposed by Alicki et al.~\cite{Alicki_QuantumEngine2023}.  
In particular, the Cooper pairs in the terminals of JJ are treated as a (macroscopic) bosonic open quantum system coupled to external electronic baths, with their dynamics governed by a Markovian Master Equation (MME)~\cite{Alicki_QuantumEngine2023}.
These tools are utilized to investigate two different designs of the Josephson pump consisting of four terminals, each in contact with an external electronic bath.
A disequilibrium between two of the baths imposes a bias, and the magnetic field is a control parameter that triggers a SQUID effect, eventually leading to a charge pumping between the other two identical baths.
We also discuss the symmetries of the pumped current under the reversal of the imposed bias and the control magnetic field.
The subtle role of the nonlinear Coulomb repulsion between different terminals is also demonstrated. 
While in \cite{Giazotto2011} a current  is pumped through a regular nanowire, in our proposals all leads are superconducting.
Unlike in the models of \cite{Thouless1983,Buttiker1994,Brouwer1998,Zhou1999}, our description of the pump does not invoke an explicit time dependence.  Why this is, and how it may be reconciled with the fact that a pump is an {\it engine} that operates cyclically, will be explored towards the end.
We now introduce the macroscopic models for a JJ.

\section{Macroscopic model of Josephson junction}
\label{sec:model}

The two best known phenomenological descriptions of macroscopic Cooper pair dynamics in a JJ are those based on Feynman's \emph{macroscopic wave function} (MWF), 
and the phenomenological \emph{resistively and capacitively shunted junction} (RCSJ) model.

In Feynman's approach, the Cooper-pair condensates in the terminals of JJ are described by a two-component MWF~\cite[Secs.~21-29]{FLP_JJ}:
\begin{align}
\vec\psi
= \begin{pmatrix}
    \psi_A \\ 
    \psi_B
\end{pmatrix}
= \begin{pmatrix}
    \sqrt{n_A} \, e^{i\phi_A} \\
    \sqrt{n_B} \, e^{i\phi_B}
\end{pmatrix}
,
\end{align}
where $n_A$ and $n_B$ are the Cooper pair populations in the two electrodes with phases $\phi_A$ and $\phi_B$, respectively.
The dynamics is governed by the Schr\"odinger equation:
\begin{align}
\pdv{t} \vec\psi = - \frac{i}{\hbar} \h H \vec\psi,
\quad
\h H =
\mqty(
+e\Delta V & \hbar K	\\
\hbar K & -e\Delta V
)
,
\end{align}
where $\Delta V$ is the voltage applied across the JJ, and $K$ is the tunneling amplitude. The current flow due to the tunneling of Cooper pairs is governed by a unitary evolution, but its impact on the charges of two electrodes is neglected.
This is because the terminals of the JJ are connected to an external battery that can provide or consume electrons in order to keep the local density of the Cooper pairs approximately constant, such that the net charge density of the corresponding material (including the fixed positive charge
density of the underlying atomic lattice) remains close to zero~\cite[Secs.~21-29]{FLP_JJ}.  
As pointed out in \cite{Alicki_QuantumEngine2023}, this makes the JJ an open system. Accordingly, any model based on the unitary Schr\"odinger equation leaves out a key part of the dynamics, and cannot explain how the JJ acts as engine, capable of outputting work in the form of a self-oscillation that causes the emission of non-thermal phonons and photons~\cite{JENKINS_2013}.

On the other hand, the RCSJ model accounts for the JJ characteristics in terms of an idealized junction described by a ``tilted washboard'' potential unbounded from below, in parallel with a resistor and a capacitor, see~\cite[Section 6.3]{Tinkham_SC} and~\cite[Secs.~4.6, 8.5]{Strogatz-book}. It was shown by Alicki et al.~\cite{Alicki_QuantumEngine2023} that an open-system description of the JJ can explain the main features of the characteristic without invoking any washboard potential and with the resistive and capacitive effects emerging from the physical parameters describing the coupling between the electrodes and the external electronic baths.  This is done in terms of the dissipation rates in the MME.

\subsection{The open system approach}
\label{sec:open}

The open system model it is a natural extension of the model proposed by Feynman, where we replace the macroscopic wave function with a \emph{macroscopic density matrix} (MDM) whose evolution in time is described by a MME~\cite{Alicki_QuantumEngine2023}. The terminals of the battery are modeled as external electron baths that are coupled to the JJ electrodes through the dissipator part of the master equation.
The condensate of Cooper pairs in a superconducting electrode occupies the ground state of a particular effective Hamiltonian, which can be represented by a single quantum harmonic oscillator. If $ \ha$ denotes the annihilation operator, the number of Cooper pairs is given by $\expect{ \ha^\dagger \ha}$, which can be increased or decreased by exchanging an electron pair with the external bath.
For a single electrode this situation is represented by an effective Hamiltonian $\h H + \h H_\text{bath} + \h V_\text{int}$, where $\h H$ describes the system, $\h H_\text{bath} = \sum_k \mathcal{E}_k \h b_k^\dagger \h b_k$ describes the electron bath that acts as the reservoir of Cooper pairs, and $\h V_\text{int}$ is the system-bath coupling:
\begin{align}
 \h V_\text{int} = \sum_{k,k'} 
\qty( g_{kk'}  \, \ha  \, \h b_k^\dagger \h b_{k'}^\dagger 
+ \bar{g}_{kk'}  \, \ha^\dagger  \, \h b_k \h b_{k'} ) .
\end{align}
The form factors $\qty{g,\bar{g}}$ contain the spectral densities of the electron bath. Note that the results of this work do not have any explicit dependence on their functional forms~\cite{Alicki_QuantumEngine2023}.
When restricted to on-shell processes, this interaction corresponds to the Andreev scattering between the Cooper pairs in the superconductor and the electrons and holes in the normal metal~\cite{Andreev1964,Sauls-Andreev}. 
The reduced density matrix of the system is governed by a master equation:
\begin{align}
\dv{\h \rho}{t}
= -\frac{i}{\hbar} [ \h H, \h \rho]
+ \mathcal{D}( \h \rho) ,
\end{align}
where $\mathcal{D}(\h \rho)$ is the dissipator due to coupling with the external electron bath.
In common experimental setups the relevant timescales for Cooper pairs (Josephson frequency, tunneling amplitudes, damping rates) are of order $10^{-12}$ s~\cite{Febvre_2010}.
For the electrons in metallic baths this timescale, as estimated with the Drude model, is usually of order $10^{-14}$ to $10^{-15}$ s. 
Such large separation of time scales implies that the bath refreshes much faster than the JJ, so that the open system dynamics is very well described in the Markovian regime.
It was argued in Ref.~\cite{Alicki_QuantumEngine2023}
that in Josephson junctions the onsite energies are proportional to the (large) number of Cooper pairs in the electrodes, where the so-called local master equation is justified.
For multiple electrodes the local form of the dissipator is written as:
\begin{align}
\mathcal{D}(\h\rho) 
=& 
\frac{1}{2} \sum_j \gamma^{\da}_j
\qty(
[ \ha_j , \h\rho \ha^{\dagger}_j ]
+ [ \ha_j \h\rho , \ha^{\dagger}_j ]
)
\nextstep
&+ \frac{1}{2}  \sum_j \gamma^{\ua}_j
\qty(
[ \ha^{\dagger}_j , \h \rho \ha_j ] + [ \ha^{\dagger}_j \h \rho ,  \ha_j ]
) 
,
\label{eq_form_of_dissipator}
\end{align}
where $\ha_j$ and $\ha_j^\dagger$ are the creation and annihilation operators for the Cooper pair condensate in $j^\text{th}$ electrode, which is put in contact with an electron bath having annihilation and creation rates $\gamma^\da_j >  \gamma^\ua_j$ [see Section 1.3 in~\cite{Alicki_QuantumSemigroups} for their explicit forms].

\begin{figure*}[!t]
\centering

\includegraphics[width=\linewidth]{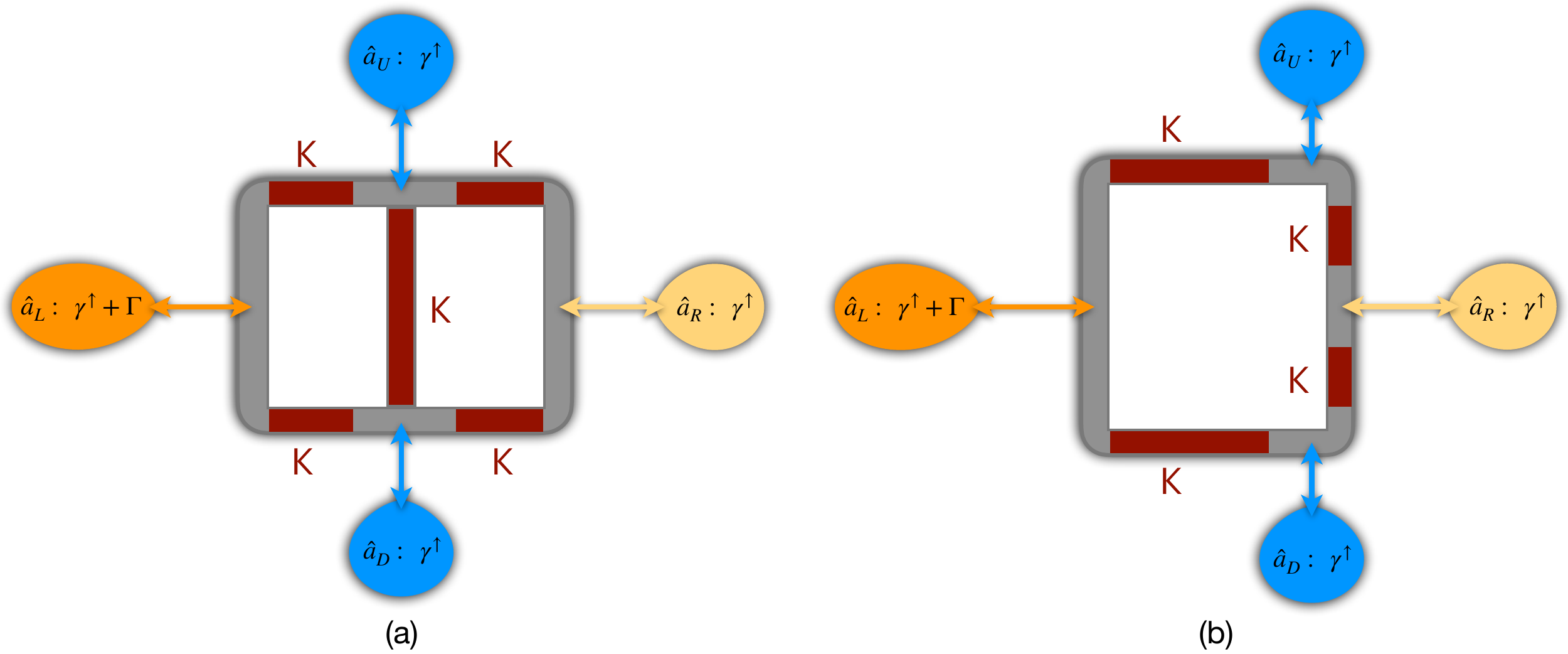}

\caption{Two geometries of the proposed Josephson pumps: (a) symmetric and (b) asymmetric. The red blocks represents the insulating barriers between the four terminals (gray blocks).
Each island is connected to its adjacent electron bath on the left-down-right-up ($L$-$D$-$R$-$U$).
$K$ is the amplitude of tunneling, and $\gamma^\ua$ is the creation rate of Cooper pairs. A bias is imposed between the $L$-$R$ modes through a disequilibrium of creation rates $\Gamma$.}

\label{fig_Setups}
\end{figure*}

\section{Josephson pumps}
\label{sec:JJ pump}

We consider two pump geometries: one that is fully symmetric and another that is asymmetric as  implemented in \cite{Giazotto2011}. These are schematically shown in \Cref{fig_Setups}.
The two geometries show different responses upon changing the polarity of the applied bias: in the symmetric setup the pumped current flips, but in the asymmetric design it is independent of the polarity.
We find that, if Coulomb interaction between electrodes is ignored, the symmetric design would allow for unlimited pumping, while the asymmetric design would not pump at all.

\Cref{fig_Setups} also introduces the notation that we will use for the parameters of the pump and for its parts.  The arrangement is composed of four terminals, labeled $LDUR$ (left, down, right, up) respectively. 
A SQUID current is imposed along the $L$-$R$ branch by a fixed difference in the electrochemical potentials of the respective baths (an applied DC bias). No bias is applied to the $D$-$U$ branch, i.e., the $D$ and $U$ baths are kept that the same electrochemical potential, as depicted with their similar color coding in \Cref{fig_Setups}.  Pumping is seen as a persistent current along the unbiased $D$-$U$ branch.  In order to break the symmetry between $D$ and $U$ we apply a magnetic flux through the circuit, as in a standard SQUID device.  The total flux of the applied magnetic field is 
$\Phi 
= \oint\! \bvec{A} \cdot \bvec{dl} $,
where $\bvec{A}$ is the vector potential corresponding to the applied magnetic field, and we follow the anticlockwise path for traversing the loop. The amplitude of tunneling from mode $j$ to $k$ is modulated as
\begin{align}
    t_{jk} = \hbar K \exp( -i\frac{2\pi}{\Phi_0} \int_j^k \bvec{A} \cdot \bvec{dl} ) , 
\end{align}
where $\Phi_0 = h/2e$ is the superconducting flux quantum, and $K$ is the bare tunneling amplitude in the absence of an external magnetic field. 
Clearly, $t_{jk} \neq t_{kj}$ in general, which leads to a SQUID effect.

In this work we consider a device with point modes, i.e., the size of terminals is much smaller than the insulating barriers. 
Assuming all JJs as physically identical, we have that
\begin{align}
  \int_L^D\!\bvec{A}\!\cdot\!\bvec{dl} 
= \int_D^R\!\bvec{A}\!\cdot\!\bvec{dl} 
= \int_R^U\!\bvec{A}\!\cdot\!\bvec{dl} 
= \int_U^L\!\bvec{A}\!\cdot\!\bvec{dl} 
= \frac{\Phi}{4}   .
\end{align}
It is now convenient to rewrite the amplitudes in the form
\begin{align}
    t_{jk} =
    \hbar K \exp(-i \frac{\delta\varphi}{(\cdot)}),
\end{align}
where $(\cdot)$ is an integer number, and $\delta\varphi$ is the total phase accumulated while threading the full outer loop:
\begin{align}
\delta\varphi 
= 2\pi \qty( \frac{\Phi}{\Phi_0} ) .
\end{align}
In this notation, the Hamiltonian of the symmetric design in \Cref{fig_Setups}(a) is given by:
\begin{align}
\h H=& 
  E \ha_{L}^{\dagger} \ha_{L} 
+ E \ha_{D}^{\dagger} \ha_{D} 
+ E \ha_{U}^{\dagger} \ha_{U} 
+ E \ha_{R}^{\dagger} \ha_{R} 
\nextline
& + \hbar K 
\big( e^{i\delta\varphi/4} \,
\ha_{L}^{\dagger} \ha_{D} + \text{h.c.} \big) + E_C ( \ha_{L}^{\dagger} \ha_{L} - \ha_{D}^{\dagger} \ha_{D} )^2 
\nextstep
& + \hbar K 
\big( e^{i\delta\varphi/4} \,
\ha_{D}^{\dagger} \ha_{R} + \text{h.c.} \big) + E_C ( \ha_{D}^{\dagger} \ha_{D} - \ha_{R}^{\dagger} \ha_{R} )^2 
\nextstep
& + \hbar K 
\big( e^{i\delta\varphi/4} \,
\ha_{R}^{\dagger} \ha_{U} + \text{h.c.} \big) + E_C ( \ha_{R}^{\dagger} \ha_{R} - \ha_{U}^{\dagger} \ha_{U} )^2 
\nextstep
& + \hbar K 
\big( e^{i\delta\varphi/4} \,
\ha_{U}^{\dagger} \ha_{L} + \text{h.c.}  \big) + E_C ( \ha_{U}^{\dagger} \ha_{U} - \ha_{L}^{\dagger} \ha_{L} )^2
\nextstep
& + \hbar K 
\big( e^{i\delta\varphi/2} \,
\ha_{D}^{\dagger} \ha_{U} + \text{h.c.} \big) + E_C ( \ha_{D}^{\dagger} \ha_{D} - \ha_{U}^{\dagger} \ha_{U} )^2 
,
\nonumber
\label{eq_HamSym}
\end{align}
where the annihilation operators $(\ha_L, \ha_D, \ha_R, \ha_U)$ identify the island in direct contact with external bath $(L,D,R,U)$.
Note that by symmetry constraints the phase difference in the branch ($D$-$U$) is twice that in all other branches constituting the outer loop. 
Similarly,
the hamiltonian for the asymmetric setup in \Cref{fig_Setups}(b) reads:
\begin{align}
\h H=& 
  E \ha_{L}^{\dagger} \ha_{L} 
+ E \ha_{D}^{\dagger} \ha_{D} 
+ E \ha_{U}^{\dagger} \ha_{U} 
+ E \ha_{R}^{\dagger} \ha_{R} 
\nextline
& + \hbar K 
\big( e^{i\delta\varphi/4} \,
\ha_{L}^{\dagger} \ha_{D} + \text{h.c.} \big) + E_C ( \ha_{L}^{\dagger} \ha_{L} - \ha_{D}^{\dagger} \ha_{D} )^2 
\nextstep
& + \hbar K 
\big( e^{i\delta\varphi/4} \,
\ha_{D}^{\dagger} \ha_{R} + \text{h.c.} \big) + E_C ( \ha_{D}^{\dagger} \ha_{D} - \ha_{R}^{\dagger} \ha_{R} )^2 
\nextstep
& + \hbar K 
\big( e^{i\delta\varphi/4} \,
\ha_{R}^{\dagger} \ha_{U} + \text{h.c.} \big) + E_C ( \ha_{R}^{\dagger} \ha_{R} - \ha_{U}^{\dagger} \ha_{U} )^2 
\nextstep
& + \hbar K 
\big( e^{i\delta\varphi/4} \,
\ha_{U}^{\dagger} \ha_{L} + \text{h.c.}  \big) + E_C ( \ha_{U}^{\dagger} \ha_{U} - \ha_{L}^{\dagger} \ha_{L} )^2 ,
\nonumber
\end{align}
which is similar to the symmetric setup expect for the junction that directly couples the $D$-$U$ branch [represented by the last line in \Cref{eq_HamSym}].

The electrostatic effects are captured through the quartic terms, as we explain below. A population disequilibrium between two modes leads to the development of an electrical double layer across the junction.
The corresponding voltage due to separated charges is proportional to the charge difference, $V_{jk} \propto \Delta n_{jk}$, and hence the electrostatic energy is given by $U_{jk} \propto V_{jk} \Delta n_{jk} = E_C \Delta n_{jk}^2$, where $E_C$ is the inverse capacitance (charging energy) of the junction.
In the Hamiltonian written above in \Cref{eq_HamSym} we have promoted the populations as number operators: $\h U_{jk} = E_C ( \h n_j - \h n_k )^2 = E_C (\ha_j^\dagger \ha_j - \ha_k^\dagger \ha_k)^2$. From now on, it will be understood that $n_j = \expect{\h  n_j} = \expect{\ha_j^\dagger \ha_j }$, and $z_{jk} = \expect{\h  z_{jk}} = \expect{\ha_j^\dagger \ha_k }$.

\begin{figure*}[!t]
\centering

\subfigure[]{\includegraphics[width=0.31\linewidth]{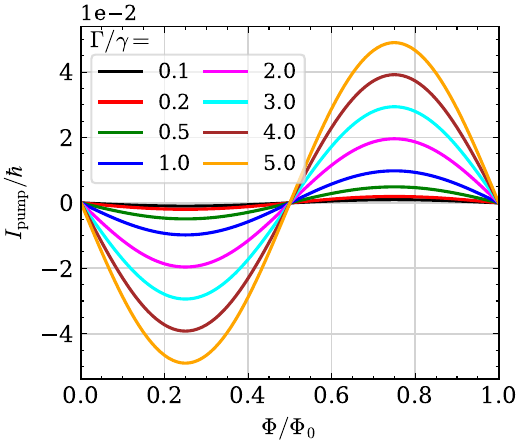}}
\hfill
\subfigure[]{\includegraphics[width=0.36\linewidth]{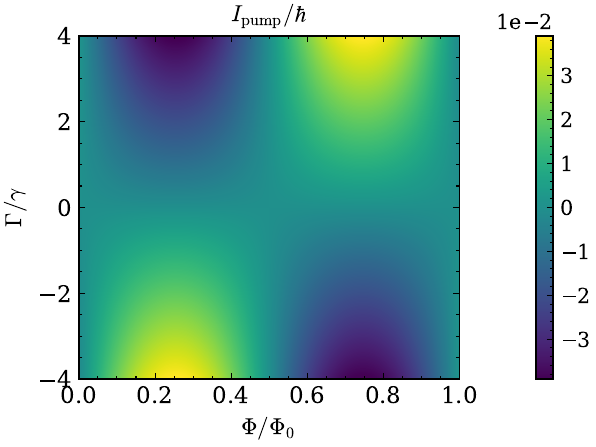}}
\hfill
\subfigure[]{\includegraphics[width=0.31\linewidth]{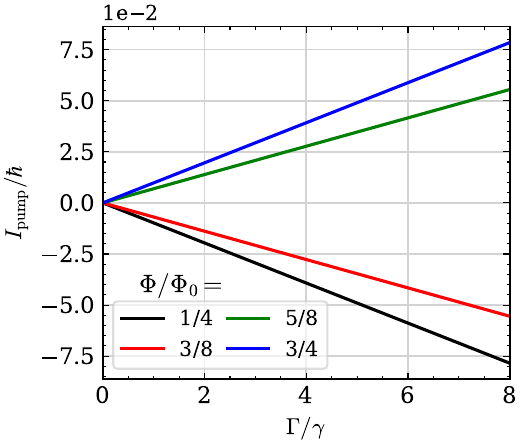}}

\subfigure[]{\includegraphics[width=0.31\linewidth]{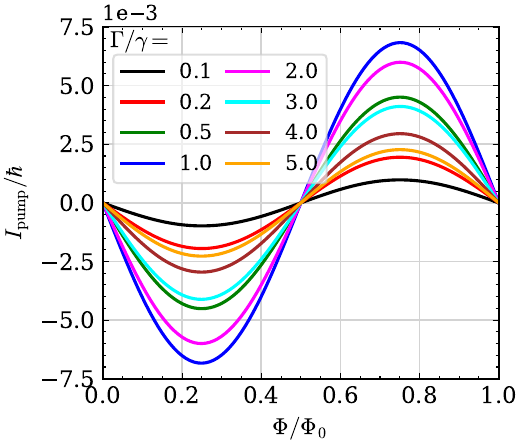}}
\hfill
\subfigure[]{\includegraphics[width=0.36\linewidth]{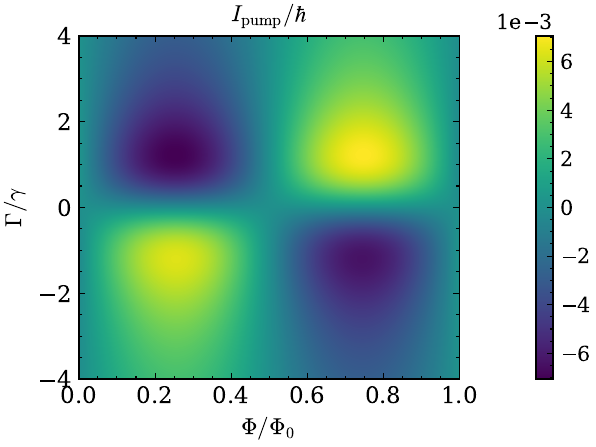}}
\hfill
\subfigure[]{\includegraphics[width=0.31\linewidth]{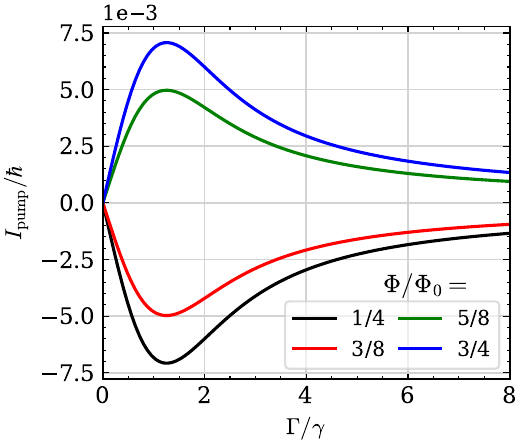}}

\caption{Current pumping in the geometry shown in \Cref{fig_Setups}(a). In this example we set $K=0.1$, $\gamma^\ua/\gamma=100$, where the constant $\gamma$ is defined in \Cref{eq:gamma}. 
$\Phi$ is the flux of the applied magnetic field shown in units of the superconducting flux quantum $\Phi_0$, and $\Gamma$ is the imposed SQUID bias.
In the top panels (a), (b), and (c), we turn off Coulomb repulsion by putting $E_C = 0$, and in the bottom panels (d), (e), and (f), we show a realistic case of $E_C/\hbar=0.1$.}

\label{fig:Pumping_Sym}
\end{figure*}

\subsection{Steady state of a general JJ network}
\label{sec:steady}

While in this work we focus on the pump setups shown in \Cref{fig_Setups}, the theoretical description shall be kept generic in order to facilitate its extension to other Josephson junction based nanodevices in future research. Hence, we start a Hamiltonian for a system composed of an arbitrary number of terminals, where all components are interacting with each other:
\begin{align}
 \h H = \sum_{j} \epsilon_{j} \ha_{j}^{\dagger} \ha_{j} + \sum_{j,k \neq j} t_{kj}  \ha_{j}^{\dagger} \ha_{k} + \sum_{j,k>j}c_{jk}(\ha_{j}^{\dagger} \ha_{j} - \ha_{k}^{\dagger} \ha_{k})^{2} .
\end{align}
The $\epsilon$'s and $c$'s are, respectively, onsite energies and inverse capacitances matrix, while the coefficients $t_{kj} = t_{jk}^*$ are the tunneling amplitudes between the respective sites. 

Let us now introduce the aforementioned macroscopic density matrix, MDM (also known as one-body reduced density matrix, or the two-point correlation matrix). For a system composed of $J$ modes the MDM is a $J\!\cross\!J$ positive definite matrix comprising all two-point correlations: $\sigma = [\sigma_{jk}] = [\expect{ \ha^\dagger_j \ha_k } ]$.
The diagonal elements are the populations: $n_j$, and the off-diagonal elements represent coherences between different modes: $z_{jk}$:
\begin{align}
\sigma 
=
\mqty(
n_{1}  & z_{12} & z_{13} & \dots & \dots & z_{1J}
\\
& n_{2} & z_{23} & \dots  & \dots & z_{2J}
\\ 
& &  \ddots & \ddots & \ddots & \vdots
\\
& & &  \ddots & \ddots & \vdots
\\
& & & & \ddots & \vdots
\\
& & & & & n_{J}
)
,
\label{eq_matrix_sigma_sturcture}
\end{align}
where we wrote out explicitly only the upper-triangular part because the MDM is hermitian, i.e., $z_{kj} = z_{jk}^*$.

The Coulomb part of the Hamiltonian has a higher-than-quadratic dependence on canonical operators, which implies that the dynamics is governed by an infinite set of coupled nonlinear equations involving ever-increasing statistical moments.
To make the MME description tractable we will work in a mean-field approximation, where we assume that the higher order correlations are negligible:
\begin{align}
    \text{Cov}(\ha_j^\dagger \ha_k, \ha_m^\dagger \ha_n) 
    =  
   \expect{ \ha_j^\dagger \ha_k \ha_m^\dagger \ha_n }
    - \expect{ \ha_j^\dagger \ha_k} \expect{ \ha_m^\dagger \ha_n } 
    \approx 0.     
\end{align}
Thus, all four-point correlations are thus replaced by products of second-order ones, $\expect{ \ha_j^\dagger \ha_k \ha_m^\dagger \ha_n } = \expect{ \ha_j^\dagger \ha_k} \expect{ \ha_m^\dagger \ha_n } = \sigma_{jk} \cdot \sigma_{mn}$, essentially giving a closed set of non-linearly coupled equations
[see \Cref{app:DeriveDynamics} for an explicit derivation].

The non-equilibrium steady-state populations and coherences are then given by
\begin{align}
n_{j}
=&  \frac{\gamma_{j}^{\ua}}{\gamma} 
+ \frac{2}{\hbar\gamma} \sum_{ m \neq j } \Im( t_{mj} z_{jm} ) ;
\label{eq:PopCoh_Generic}
\nextline
z_{jk}
=& \frac{ 
t_{jk} \Delta n_{jk}
+ \sum_{ m \neq j,k} \qty( t_{mk} z_{jm}  - t_{jm} z_{mk} ) 
}
{
i\hbar\gamma  
+ \Delta\epsilon_{jk}
+ 2 \sum_m \qty( c_{jm}  \Delta n_{jm} +  c_{mk} \Delta n_{mk} )
}
,
\nonumber
\end{align}
where $\Delta n_{jk} = n_j - n_k$, and for simplicity we assumed that the difference between pumping and damping rates to be the same for all baths, so that we may work with a single parameter:
\begin{align}
\label{eq:gamma}
    \gamma= \gamma_j^\da - \gamma_j^\ua, 
    \hspace{5mm} \forall j.
\end{align}
The steady state conditions in \Cref{eq:PopCoh_Generic} are numerically solved through a self-consistent fixed-point iteration for the populations and the coherences. 
The initial guesses are generated randomly, and we find for a wide range of parameters that the iterative algorithm converges to the same final fixed-points irrespective of these random inputs.
In this work we set the acceptable error tolerance as $ \sum_{j} | n_{j}^{(1)} - n_{j}^{(0)} | < 10^{-8}$, where $n^{(1)}$ is the next iterative guess based on the current solution $n^{(0)}$.

\subsection{Pumping in the two geometries}
\label{sec:geometries}

We now use \Cref{eq:PopCoh_Generic} to obtain the non-equilibrium steady state for the proposed setups in \Cref{fig:Pumping_Sym}.  
The charge current flowing out of the $j^\text{th}$ mode is simply the part of its population that is being dissipated into the external bath. Explicitly calculated in \Cref{eq_DissipatorPart} in \Cref{app:DeriveDynamics}, this is given by:
\begin{align}
I_j = \Tr\qty( \ha^\dagger_j \ha_j \mathcal{D}(\h\rho) ) 
=  -\gamma n_j + \gamma_j^\ua .
\end{align}
Accordingly, the pumped current is:
\begin{align}
I_\text{pump} = I_D - I_U = - \gamma \Delta n_{DU} .
\end{align}
In \Cref{fig:Pumping_Sym} we consider the symmetric design shown in \Cref{fig_Setups}(a).
For representative purposes, in the top panels we have turned off the Coulomb interaction.
It can be seen in panel (a) that the amplitude of $I_\text{pump}$ increases monotonically with the bias. The open contours in the heat plot of panel (b) confirm the same. 
This clearly contradicts the experimental results showing the amplitude of $I_\text{pump}$ as bounded from above~\cite{Giazotto2011}.
The bottom panels show the more realistic description with a nonzero Coulomb repulsion.
Panel (d) shows that the corresponding amplitude first increases with the bias, but after some level it starts decreasing and eventually decays to zero. 
Panel (e) shows that the heatmap contour is closed, which implies that the pumping is bounded from above.
For an explicit demonstration of the significance of Coulombic terms, we also show the pumped current as a function of the applied bias for some chosen values of the magnetic flux.
It can now be clearly seen that the current profile is unrestricted for $E_C = 0$ in panel (c), and it agrees with experimental findings only after the inclusion of the electrostatic terms as shown in panel (f).
Note also the smaller values of $I_\text{pump}$ in the bottom panels, which is expected because the Coulomb repulsion suppresses the tunneling of Cooper pairs by making it energetically costly to vary local densities from their equilibrium.

\begin{figure*}[!t]
\centering

\subfigure[]{\includegraphics[width=0.31\linewidth]{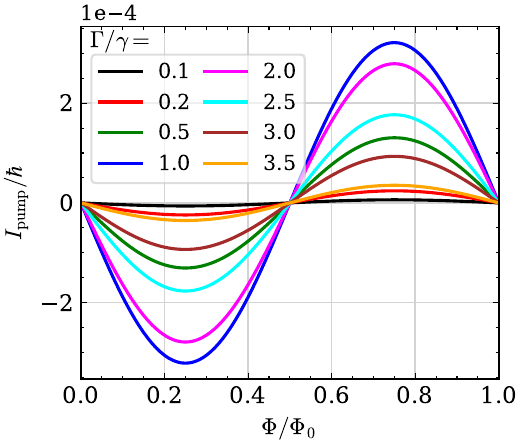}}
\hfill
\subfigure[]{\includegraphics[width=0.36\linewidth]{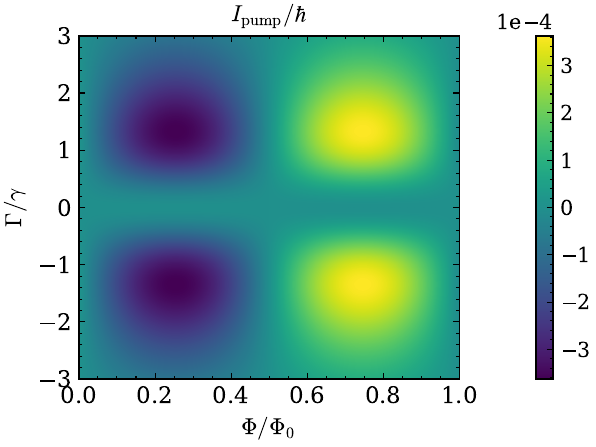}}
\hfill
\subfigure[]{\includegraphics[width=0.31\linewidth]{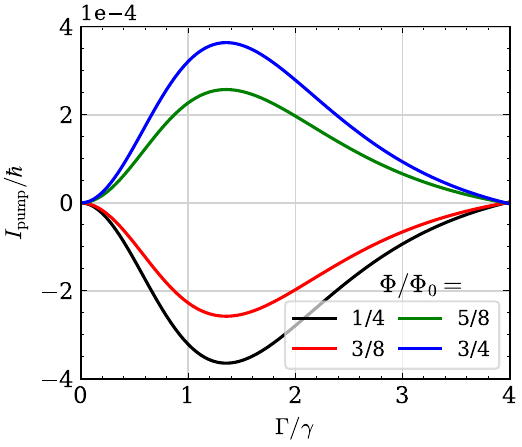}}

\caption{Current pumping in the geometry shown in \Cref{fig_Setups}(b). In this example we set $K=0.1$, $E_C/\hbar=0.1$, $\gamma^\ua/\gamma=100$, where the constant $\gamma$ is defined in \Cref{eq:gamma}. 
$\Phi$ is the flux of the applied magnetic field shown in units of the superconducting flux quantum $\Phi_0$, $I_\text{pump}$ is the pumped current, and $\Gamma$ is the imposed SQUID bias.}

\label{fig:Pumping_Asym}
\end{figure*}

We now consider the asymmetric design of \Cref{fig_Setups}(b), which is qualitatively similar to the experiment realized in \cite{Giazotto2011}, except that in our case all of the leads are superconducting. The pumping in this geometry is shown in \Cref{fig:Pumping_Asym}.
Note that the charge is now pumped sequentially through multiple modes, while in the symmetric setup in \Cref{fig_Setups}(a) there was an additional channel for direct tunneling between $D-U$.  
Hence, the asymmetric setup therefore is subject to a stronger Coulomb blockade, which explains why $I_\text{pump}$ is an order of magnitude smaller as compared to the symmetric setup.
The pumping is flipped under a reversal of the magnetic flux, but it stays the same under  a reversal of the applied bias. This in qualitative agreement with the experimental findings reported in \cite{Giazotto2011}.
Thus, we find that by eliminating the normal wire of~\cite{Giazotto2011}, the qualitative picture of charge pumping remains the same, but the resistive heat losses associated with charge transport could be reduced.
Note that we do not see any charge pumping for this second geometry if Coulombic terms are not included in the Hamiltonian, as we will show below.

\begin{figure}[!b]
    \centering
    \includegraphics[width=\linewidth]{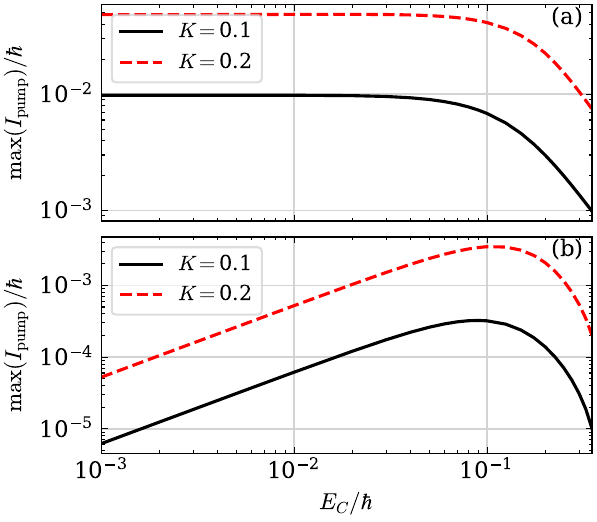}
    \caption{Effects of Josephson junction capacitance on the pumped current. 
    In this example we set $\gamma^\ua/\gamma=100$, where the constant $\gamma$ is defined in \Cref{eq:gamma}. $K$ is the tunneling constant, $\Gamma/\hbar = 1$ is the imposed SQUID bias, and $E_C$ is the inverse capacitive constant of the junctions. $I_\text{pump}$ is the pumped current, which has been maximized over all values of the magnetic flux. 
    Panels (a) and (b) represent the symmetric and the asymmetric designs presented in \Cref{fig_Setups}.}
    \label{fig:FIG_Pumping_w_Coulomb}
\end{figure}

In \Cref{fig:FIG_Pumping_w_Coulomb} we show the impact of the strength of Coulomb repulsion on the amplitude of pumped current. In the symmetric design of \Cref{fig_Setups}(a), the amplitude is largest for $E_C = 0$ and it decreases monotonically for increasing values of the Coulomb strength. However, for the asymmetric design of \Cref{fig_Setups}(b), the pumping vanishes in both the $E_C \to 0$ and the $E_C \to \infty$ regime, while achieving a maximum in some intermediate region.  It is therefore clear that the electrostatic interaction between terminals must be taken into account to obtain a physically reasonable description of the mechanism of charge pumping in these devices.

\subsection{Hidden cycles} 
\label{sec:cycles}

\begin{figure*}[!t]
\centering
 
\includegraphics[width=\linewidth]{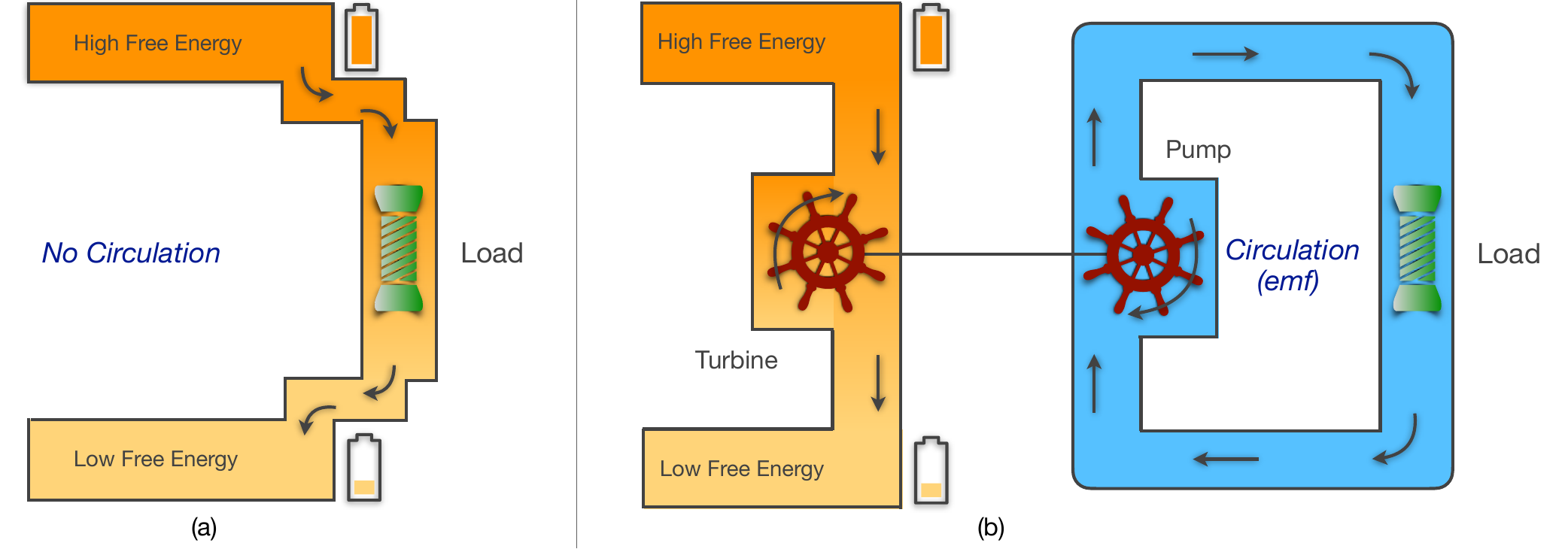}

\caption{(a) A passive device that consumes free energy from an external source. (b) An active device (engine) that uses some of the free energy consumed from an external source to perform sustained work. These illustrations are adapted from Ref.~\cite{LEC2021}.}

\label{fig:Elblag}
\end{figure*}

A charge pump is an {\it engine}, i.e., an open system that can perform work cyclically, at the expense of an external disequilibrium (in this case, the DC bias applied along the $L$-$R$ branch in \Cref{fig:Pumping_Sym}).  For a general discussion of this point of view, see~\cite{Engines2021}. \Cref{fig:Elblag} shows a general scheme for two types of irreversible processes.  Panel (a) represents a passive process that consumes free energy from an external source.  Such a process cannot maintain any circulation of flow.  The panel (b) represents an active process that uses part of the free energy consumed to perform a sustained work, represented there by the turning of a turbine in order to pump the active flow in the circuit on the right.  In an electrical device, the integral per unit charge of the active, non-conservative force that drives this circulation is called the {\it electromotive force} (emf).  Such an active process corresponds to an engine in the sense defined above.

In the theoretical treatments of adiabatic quantum pumps of~\cite{Thouless1983, Buttiker1994, Brouwer1998, Zhou1999}, pumping is obtained from the time dependence of a potential or scattering amplitude.  Ref.~\cite{Dreon2022} considers a quantum pump in which the time dependence is generated endogenously (i.e., as a {\it self-oscillation}) as conceptualized in Ref.~\cite{JENKINS_2013}, rather than imposed as an external periodic driving. In both cases the pumping can be interpreted in terms of an engine {\it cycle}. In Giazzotto et al.'s modeling of their Josephson pump, this cycle appears as the dependence upon the JJ phase ($\dot \phi = 2eV/\hbar$) of the amplitude for Andreev scattering \cite[Supplementary Information]{Giazotto2011}. Thus, in terms of the scheme of \Cref{fig:Elblag}(b), we may say that the scheme of \cite{Giazotto2011} uses the JJ phase as the ``turbine'' that drives the pump current.

We have argued that our description, which is based on a master equation for the JJ treated as an open quantum system, can clarify the thermodynamic analysis of the Josephson pump.  But it also raises a puzzle: Where are the engine cycle and the associated ``turbine'' in this description?  This is an important point, because at the first glance it could appear from our own treatment that the process described is analogous to the Hall effect, in which a voltage applied longitudinally along a conductor produces, by the action on an external magnetic field, a voltage in the transverse direction.  However, in the scheme of \Cref{fig:Elblag} the Hall effect is not an instance of charge pumping, since there the generation of a transverse voltage results from a classical Lorentz force, which does no work on the charges.  Indeed, it has been shown that most of the power that drives the lateral current between the edges of the Hall bar comes directly from the applied longitudinal current, rather than from the perpendicular current associated with the Hall voltage \cite{Hall2021}.  Moreover, the transverse voltage in a Hall bar is not an emf in the sense of \Cref{fig:Elblag}(b), since it cannot drive charges along a closed path.

It is therefore clear that the setup of \Cref{fig_Setups} is qualitatively different from the Hall effect, as here the charges experience no classical Lorentz force, and the role of the magnetic field is only to break the $D$-$U$ symmetry by introducing a relative phase in the amplitudes of quantum tunneling between different terminals. But it may still seem puzzling that we should have been able to describe the active pumping of \cite{Giazotto2011} in terms of equations with no explicit time dependence or cycles.

In the model of the JJ presented in \cite{Alicki_QuantumEngine2023}, the external bias sustains the coherences $[z_{jk}]$ against decay from the interaction with the baths.  One of the key results of \cite{Alicki_QuantumEngine2023} is that the equations of motion for the JJ engine have ``hidden attractor'' cycles around fixed points with non-zero coherence.  It is these hidden cycles that explain how the JJ acts as an engine, generating work in the form of nonthermal sound and electromagnetic waves at the Josephson frequency ($\Omega = 2eV / \hbar$) and its lower harmonics.  In the model of the Josephson pump presented in this article, it is the sustained coherences $[z_{jk}]$ that act as cyclical ``turbine'' variables, which drive the pumped current.

The MME formalism is averaged over the period of these hidden attractors, so that the time dependence associated with the pumping cycle does not appear explicitly in our equations.  We expect, however, that the active pumping of charge from $D$ to $U$ that we obtained in our calculations reflects the modulation of the tunneling rates by the coherences $[z_{jk}]$ along the hidden attractors identified in \cite{Alicki_QuantumEngine2023}.  The details of this, as well as the broader question of how non-equilibrium steady states more generally could be understood as time averages over hidden cycles, call for further investigation. We regard this as one of the most interesting and pressing questions raised by this work.

\section{Discussion}
\label{sec:discussion}

By treating the Cooper pairs in the JJ as a bosonic open quantum system, governed by a MME, we developed a simple and robust model for studying the dynamics of nanodevices based on the JJ. Using these theoretical tools we explored charge pumpings similar to the one realized in \cite{Giazotto2011}.  In the setup of \cite{Giazotto2011} the current is pumped through a regular nanowire, which leads to resistive heat losses.
In the current proposal this is avoided by using only superconducting leads. The suppression of electrical resistance gives negligible heat losses, which could allow for a coherent operation over longer runtimes.  We found that if the Coulomb interaction between the terminals is included in the model, via nonlinear terms in the Hamiltonian, the results obtained are in good qualitative agreement with available experimental data.

Although the steady state that we obtained from the MME has no explicit time dependence, the presence of ``hidden attractor'' cycles in the JJ equations of motion obtained in \cite{Alicki_QuantumEngine2023} suggests that our results are consistent with the presence of an engine cycle like that seen in other active systems.  This may be of broader interest in light of the fact that major theoretical questions still surround the dynamics responsible for the generation of an emf in batteries, solar cells, thermoelectric generators, fuel cells, and other active devices of great practical importance.  It has been proposed that these devices have hidden cycles associated with internal mechanical degrees of freedom; see, e.g., \cite{Alicki2016, solarcell2017, solarcell2019, battery2021}.  By explicitly showing how the JJ can pump a steady current and how this is connected with a hidden internal cycle, this work could help to shed light on the broader problem of the dynamical generation of an emf in active systems.

We regard it as promising that a simple open-system model of the Josephson quantum pump, as the one presented here, is sufficient to qualitatively account for all the features reported by experimentalists.  We hope that this will help to clarify the sense in which the AC Josephson effect should be understood as a non-unitary, thermodynamically irreversible dynamics, as previously advocated in \cite{Alicki_QuantumEngine2023}. Although still rather phenomenological, the model developed in this work promises to expand the footprint of MME to the dynamics of superconducting nanodevices. 
In view of the fact that the simple model presented here succeeds in capturing the basic mechanism of Josephson pumps, we consider it imperative to explore how it may be adapted and extended to other devices, and also to better ground such a model in terms of the relevant microphysics.  One broad problem in which this modeling can be relevant concerns the relation between non-equilibrium steady states and the time-averaging over hidden cycles in active systems such as the Josephson pump.

\begin{acknowledgments}
We thank Robert Alicki for discussions.  
A.K. was supported by the QuantERA II project ``ExTRaQT'' grant No: 2021/03/Y/ST2/00178 that has received funding from EU’s Horizon 2020,
and by NCN SONATA-BIS grant No: 2017/26/E/ST2/01008.  
L.C. was supported by the Polish National Science Centre grant OPUS-21 (No: 2021/41/B/ST2/03207). 
M.H. acknowledges support by the IRA Programme, project no. FENG.02.01-IP.05-0006/23, financed by the FENG program 2021-2027, Priority FENG.02, Measure FENG.02.01., with the support of the FNP.
A.J. was supported by the Vicerrectorate of Research of the University of Costa Rica, under project no.\ 112-C1-716, ``Apoyo al Laboratorio de Física Teórica y Computacional de la Escuela de Física''.
The authors gratefully thank the anonymous referee for their valuable comments, which added to the clarity and completeness of this article.
\end{acknowledgments}

\onecolumngrid
\appendix

\section{Markovian dynamics for a JJ network}
\label{app:DeriveDynamics}

The Hamiltonian considered in this work is of the form
\begin{align}
\h  H = \sum_{j} \epsilon_{j} \ha_{j}^{\dagger} \ha_{j} + \sum_{j,k \neq j} t_{kj} \ha_{j}^{\dagger} \ha_{k} + \sum_{j,k>j} c_{jk}(\ha_{j}^{\dagger} \ha_{j} - \ha_{k}^{\dagger} \ha_{k})^{2} ,
\end{align}
where $c_{jk} = c_{kj}$ are elements of the inverse capacitance matrix, and $t_{kj}=t_{jk}^*$ are tunneling amplitudes between modes $k$ and $j$. The Cooper pair condensate in a superconducting electrode is modelled as a single quantum harmonic oscillator satisfying the following bosonic commutation relations:
\begin{align}
    \big[ \ha_j, \ha_k \big] = 0,
\quad
    \big[ \ha_j^\dagger, \ha_k^\dagger \big] = 0,
\quad
    \big[ \ha_j, \ha_k^\dagger \big] = \delta_{jk}.
\end{align}
The open dynamics of the system is governed  by the local MME:
\begin{align}
\dv{\h \rho}{t} = -\frac{i}{\hbar} \comm{\h H}{\h \rho} 
+ \mathcal{D}( \h \rho) , 
\quad
: \qty{
\mathcal{D}(\h \rho) =
\frac{1}{2}\sum_{j}\gamma_{j}^{\da}
\qty(
[ \ha_{j} , \h \rho \ha_{j}^{\dagger} ]
+ [ \ha_{j} \h \rho , \ha_{j}^{\dagger} ]
)
+ \frac{1}{2}\sum_{j}\gamma_{j}^{\ua}
\qty(
[ \ha_{j}^{\dagger} , \h \rho \ha_{j} ] + [ \ha_{j}^{\dagger} \h \rho , \ha_{j} ]
)
} ,
\end{align}
where $\h H$ is the system Hamiltonian and $\mathcal{D}(\h \rho)$ is the dissipator due to the external bath.
Accordingly, the time evolution of the MDM is given by:
\begin{align}
    \dv{t} \sigma_{jk} 
=& \Tr\qty( \ha^\dagger_j \ha_k \dv{\h \rho}{t} ) 
\nextstep
=&  -\frac{i}{\hbar} \Tr\qty( \ha^\dagger_j \ha_k  [\h H,\h \rho] ) + \Tr\qty( a^\dagger_j \ha_k \mathcal{D}(\h \rho) ) 
\nextstep
=&  -\frac{i}{\hbar} \Tr\qty( \ha^\dagger_j \ha_k \h H \h  \rho - \ha^\dagger_j \ha_k \h \rho \h H ) + \Tr\qty(  \ha^\dagger_j \ha_k \mathcal{D}(\h \rho) ) 
\nextstep
=&  -\frac{i}{\hbar} \Tr\qty( \ha^\dagger_j \ha_k \h  H \h \rho - \h H \h a^\dagger_j \ha_k  \h \rho ) + \Tr\qty(  \ha^\dagger_j \ha_k \mathcal{D}( \h \rho) ) 
\nextstep
=&  -\frac{i}{\hbar} \ev{ [\ha^\dagger_j \ha_k, \h   H] } + \Tr\qty( \ha^\dagger_j \ha_k \mathcal{D}(\h \rho) ) .
\end{align}
Going forward, the following identity will be quite useful:
\begin{align}
\qty[ \ha_{j}^{\dagger} \ha_{k} , \ha_{m}^{\dagger} \ha_{m} ]  
&= \ha_{j}^{\dagger} \ha_{k}  \ha_{m}^{\dagger} \ha_{m} - \ha_{m}^{\dagger} \ha_{m} \ha_{j}^{\dagger} \ha_{k}
\nextstep
&= \ha_{j}^{\dagger} (\delta_{km} + \ha_{m}^{\dagger} \ha_{k} ) \ha_{m} - \ha_{m}^{\dagger} ( \delta_{jm} + \ha_{j}^{\dagger}  \ha_{m} ) \ha_{k} 
\nextstep
&= \ha_{j}^{\dagger} \ha_{m} \delta_{km} - \ha_{m}^{\dagger} \ha_{k} \delta_{jm}
\nextstep
&\equiv ( \delta_{km} - \delta_{jm} ) \ha_j^\dagger  \ha_k .    
\end{align}
Due to the onsite energy part we have
\begin{align}
\sum_{m} \epsilon_{m} \ev{ 
[ \ha_j^\dagger \ha_k , \ha_m^\dagger \ha_m ]
}
=  \sum_{m} \epsilon_{m}  \ev{ \qty( \delta_{km} - \delta_{jm} )  \ha_j^\dagger  \ha_k }
= - \Delta \epsilon_{jk} \sigma_{jk}
,
\end{align}
where $\Delta (\cdot)_{jk} = (\cdot)_j - (\cdot)_k$.
The tunnelings between different modes contribute to:
\begin{align}
&
\sum_{m,n} t_{nm}
 \ev{ 
[ \ha_j^\dagger \ha_k ,  \ha_m^\dagger \ha_n ]
}
\nextstep
&=   \sum_{m,n} t_{nm} \ev{ 
\ha_j^\dagger \ha_k  \ha_m^\dagger \ha_n - \ha_m^\dagger \ha_n  \ha_j^\dagger \ha_k
}
\nextstep
&=  \sum_{m,n} t_{nm} \ev{ 
\ha_j^\dagger ( \delta_{km} + \ha_m^\dagger \ha_k )  \ha_n - \ha_m^\dagger ( \delta_{jn} + \ha_j^\dagger \ha_n) \ha_k
}
\nextstep
&=   \sum_{ n } t_{nk} \sigma_{jn} - 
 \sum_{m} t_{jm} \sigma_{mk}
\nextstep
&=   \sum_{ m } \qty( t_{mk} \sigma_{jm}  - t_{jm} \sigma_{mk} )
.
\end{align}

Note that $c_{mm} = 0$, and hence for the electrostatic part we have
\begin{align}
&   \sum_{m,n>m} c_{mn} \ev{ 
[ \ha_j^\dagger \ha_k ,
 (\ha_m^\dagger \ha_m - \ha_n^\dagger \ha_n)^2 ]
}
\nextstep
&=  \frac{1}{2} \sum_{m,n} c_{mn} \ev{ 
[ \ha_j^\dagger \ha_k ,
 (\ha_m^\dagger \ha_m - \ha_n^\dagger \ha_n)^2 ]
}
\nextstep
&=   \frac{1}{2}  \sum_{m,n} c_{mn} 
\bigg(
\ev{ 
[ \ha_j^\dagger \ha_k ,
 (\ha_m^\dagger \ha_m - \ha_n^\dagger \ha_n )] (\ha_m^\dagger \ha_m - \ha_n^\dagger \ha_n) }
+ \ev{
(\ha_m^\dagger \ha_m - \ha_n^\dagger \ha_n) 
[ \ha_j^\dagger \ha_k,
 (\ha_m^\dagger \ha_m - \ha_n^\dagger \ha_n) ]
} 
\bigg)
\nextstep
&=   \frac{1}{2}  \sum_{m,n} c_{mn} 
\qty( \delta_{km} - \delta_{jm} - \delta_{kn} + \delta_{jn} )
\bigg(
\ev{ 
 \ha_j^\dagger  \ha_k  (\ha_m^\dagger \ha_m - \ha_n^\dagger \ha_n) }
+ \ev{
(\ha_m^\dagger \ha_m - \ha_n^\dagger \ha_n)  \ha_j^\dagger  \ha_k 
}
\bigg) 
.
\end{align}
This involves fourth-order moments, and their dynamics will involve sixth-order moments, and so on, making the set of equations open. In order to close the set we shall work in a mean-field approximation (MFA) that neglects the higher-order correlations by assuming
\begin{align}
\text{Cov}(\ha_j^\dagger \ha_k, \ha_m^\dagger \ha_n) = \expect{\ha_j^\dagger \ha_k \ha_m^\dagger \ha_n} - \expect{\ha_j^\dagger \ha_k} \expect{\ha_m^\dagger \ha_n} \approx 0 .    
\end{align}
Accordingly, all fourth-order moments are replaced by a product of second-order ones:
\begin{align}
	\expect{\ha_j^\dagger \ha_k \ha_m^\dagger \ha_n} = \expect{\ha_j^\dagger \ha_k} \expect{\ha_m^\dagger \ha_n} = \sigma_{jk} \cdot \sigma_{mn} .
\end{align}
Thus, we obtain
\begin{align}
&  \sum_{m,n>m} c_{mn} \ev{ 
[ \ha_j^\dagger \ha_k ,
 (\ha_m^\dagger \ha_m - \ha_n^\dagger \ha_n)^2 ]
}
\nextstep
&= 
 \frac{1}{2}  \sum_{m,n} c_{mn} ( \delta_{km} - \delta_{jm} - \delta_{kn} + \delta_{jn} )
\Big(
\sigma_{jk} (n_m-n_n) + (n_m-n_n) \sigma_{jk}
\Big)
\nextstep
&= \sigma_{jk} \sum_{m,n} c_{mn} 
( \delta_{km} - \delta_{jm} - \delta_{kn} + \delta_{jn} ) \Delta n_{mn}
\nextstep
&=  \sigma_{jk} \qty( 
  \sum_{n} c_{kn} \Delta n_{kn} 
- \sum_{n} c_{jn} \Delta n_{jn} 
- \sum_{m} c_{mk} \Delta n_{mk} 
+ \sum_{m} c_{mj} \Delta n_{mj} 
)
\nextstep
&=  \sigma_{jk} \qty( 
- \sum_{m} c_{mk} \Delta n_{mk} 
- \sum_{m} c_{jm} \Delta n_{jm} 
- \sum_{m} c_{mk} \Delta n_{mk} 
- \sum_{m} c_{jm} \Delta n_{jm} 
)
\nextstep
&= -2 \sigma_{jk} \sum_{m} \qty( c_{jm}  \Delta n_{jm} +  c_{mk} \Delta n_{mk}
) .
\end{align}
We now calculate the dissipator part. In the Markovian local master equation, its contribution to evolution of the two-point correlation matrix is:
\begin{align}
\Tr\qty(  \ha^\dagger_j \ha_k \mathcal{D}(\h \rho) )
= 
\frac{1}{2}\sum_{m}\gamma_{m}^{\da}
\Tr\qty( \ha^\dagger_j \ha_k
[\ha_{m},\h \rho \ha_{m}^{\dagger}] + \ha^\dagger_j \ha_k[\ha_{m}\h \rho,\ha_{m}^{\dagger}]
)
+
\frac{1}{2}\sum_{m}\gamma_{m}^{\ua}
\Tr\qty( \ha^\dagger_j \ha_k
[\ha_{m}^{\dagger},\h \rho \ha_{m}] + \ha^\dagger_j \ha_k[\ha_{m}^{\dagger}\h \rho,\ha_{m}]
) .
\end{align}
The decomposition part is
\begin{align}
&
\frac{1}{2}\sum_{m}\gamma_{m}^{\da}
\Tr\qty(  \ha^\dagger_j \ha_k
[ \ha_{m}, \h \rho  \ha_{m}^{\dagger}] +  \ha^\dagger_j \ha_k[ \ha_{m} \h \rho, \ha_{m}^{\dagger}]
)
\nextstep
&=
\frac{1}{2}\sum_{m}\gamma_{m}^{\da}\Tr\qty(2 \ha_{j}^{\dagger} \ha_{k} \ha_{m} \h \rho  \ha_{m}^{\dagger}- \ha_{j}^{\dagger} \ha_{k} \h \rho  \ha_{m}^{\dagger} \ha_{m}- \ha_{j}^{\dagger} \ha_{k} \ha_{m}^{\dagger} \ha_{m} \h \rho)
\nextstep
&=
\frac{1}{2}\sum_{m}\gamma_{m}^{\da}\Tr\qty(2 \ha_{m}^{\dagger} \ha_{j}^{\dagger} \ha_{k} \ha_{m} \h \rho- \ha_{m}^{\dagger} \ha_{m} \ha_{j}^{\dagger} \ha_{k} \h \rho - \ha_{j}^{\dagger} \ha_{k} \ha_{m}^{\dagger} \ha_{m} \h \rho)
\nextstep
&=
\frac{1}{2}\sum_{m}\gamma_{m}^{\da}\ev{2 \ha_{m}^{\dagger} \ha_{j}^{\dagger} \ha_{k} \ha_{m}- \ha_{m}^{\dagger} \ha_{m} \ha_{j}^{\dagger} \ha_{k}- \ha_{j}^{\dagger} \ha_{k} \ha_{m}^{\dagger} \ha_{m} }
\nextstep
&=
\frac{1}{2}\sum_{m}\gamma_{m}^{\da} \ev{ 2 \ha_{m}^{\dagger} \ha_{j}^{\dagger} \ha_{k} \ha_{m}- \ha_{j}^{\dagger} \ha_{k} \ha_{m}^{\dagger} \ha_{m}+(\delta_{km}-\delta_{jm}) \ha_{j}^{\dagger} \ha_{k}- \ha_{j}^{\dagger} \ha_{k} \ha_{m}^{\dagger} \ha_{m} }
\nextstep
&=
\frac{1}{2}(\gamma_{k}^{\da}-\gamma_{j}^{\da}) \sigma_{jk} 
+ \sum_{m}\gamma_{m}^{\da}\ev{ \ha_{m}^{\dagger} \ha_{j}^{\dagger} \ha_{k} \ha_{m}- \ha_{j}^{\dagger} \ha_{k} \ha_{m}^{\dagger} \ha_{m}} 
\nextstep
&=
\frac{1}{2}(\gamma_{k}^{\da}-\gamma_{j}^{\da})\sigma_{jk} 
+ \sum_{m}\gamma_{m}^{\da}\ev{ \ha_{j}^{\dagger} \ha_{m}^{\dagger} \ha_{m} \ha_{k}- \ha_{j}^{\dagger}(\delta_{km}+ \ha_{m}^{\dagger} \ha_{k}) \ha_{m}}
\nextstep
&=
\frac{1}{2}(\gamma_{k}^{\da}-\gamma_{j}^{\da})\sigma_{jk} - \gamma_{k}^{\da} \sigma_{jk}
\nextstep
&=
-\frac{1}{2}(\gamma_{j}^{\da}+\gamma_{k}^{\da}) \sigma_{jk} 
,
\end{align}
and the creation part is
\begin{align}
&
\frac{1}{2}\sum_{m}\gamma_{m}^{\ua}
\Tr\qty( \ha^\dagger_j \ha_k
[\ha_{m}^{\dagger},\h \rho \ha_{m}] + \ha^\dagger_j \ha_k[\ha_{m}^{\dagger}\h \rho,\ha_{m}]
)
\nextstep
&=
\frac{1}{2}\sum_{m}\gamma_{m}^{\ua}\Tr\qty(2\ha_{j}^{\dagger}\ha_{k} \ha_{m}^{\dagger} \h \rho \ha_{m}-\ha_{j}^{\dagger} \ha_{k} \h \rho \ha_{m} \ha_{m}^{\dagger}- \ha_{j}^{\dagger} \ha_{k} \ha_{m} \ha_{m}^{\dagger} \h \rho)
\nextstep
&=
\frac{1}{2}\sum_{m}\gamma_{m}^{\ua}\Tr\qty(2 \ha_{m} \ha_{j}^{\dagger} \ha_{k} \ha_{m}^{\dagger} \h \rho- \ha_{m} \ha_{m}^{\dagger} \ha_{j}^{\dagger} \ha_{k} \h \rho- \ha_{j}^{\dagger} \ha_{k} \ha_{m} \ha_{m}^{\dagger} \h \rho)
\nextstep
&=
\frac{1}{2}\sum_{m}\gamma_{m}^{\ua}\ev{2 \ha_{m} \ha_{j}^{\dagger} \ha_{k} \ha_{m}^{\dagger}- \ha_{m} \ha_{m}^{\dagger}a_{j}^{\dagger} \ha_{k}- \ha_{j}^{\dagger} \ha_{k} \ha_{m} \ha_{m}^{\dagger}}\nextstep
&=
\frac{1}{2}\sum_{m}\gamma_{m}^{\ua}\ev{2 \ha_{m} \ha_{j}^{\dagger} \ha_{k} \ha_{m}^{\dagger}- \ha_{m} \ha_{m}^{\dagger} \ha_{j}^{\dagger} \ha_{k}- \ha_{m}^{\dagger} \ha_{m} \ha_{j}^{\dagger} \ha_{k}-(\delta_{km}-\delta_{jm}) \ha_{j}^{\dagger} \ha_{k}}
\nextstep
&=
-\frac{1}{2}(\gamma_{k}^{\ua}-\gamma_{j}^{\ua})\sigma_{jk} 
+ \sum_{m}\gamma_{m}^{\ua}\ev{ \ha_{m} \ha_{j}^{\dagger} \ha_{k} \ha_{m}^{\dagger}- \ha_{m} \ha_{m}^{\dagger} \ha_{j}^{\dagger} \ha_{k}}
\nextstep
&=
-\frac{1}{2}(\gamma_{k}^{\ua}-\gamma_{j}^{\ua})\sigma_{jk}
+
\sum_{m}\gamma_{m}^{\ua}\ev{ \ha_{m} \ha_{j}^{\dagger}(\delta_{km} +  \ha_{m}^{\dagger} \ha_{k})- \ha_{m} \ha_{m}^{\dagger} \ha_{j}^{\dagger} \ha_{k}}
\nextstep
&=
-\frac{1}{2}(\gamma_{k}^{\ua}-\gamma_{j}^{\ua})\sigma_{jk}
+ \gamma_{k}^{\ua}\ev{ \ha_{k} \ha_{j}^{\dagger}}
\nextstep
&=
-\frac{1}{2}(\gamma_{k}^{\ua}-\gamma_{j}^{\ua})\sigma_{jk}
+ \gamma_{k}^{\ua}\ev{ \delta_{jk} +  \ha_{j}^{\dagger} \ha_{k} }
\nextstep
&=
\frac{1}{2}(\gamma_{j}^{\ua} + \gamma_{k}^{\ua})\sigma_{jk} + \gamma_{k}^{\ua}\delta_{jk}
.
\end{align}
This implies that total dissipation is
\begin{align}
\Tr\qty( \ha^\dagger_j \ha_k \mathcal{D}(\h \rho) )
&= 
\gamma_{k}^{\ua}\delta_{jk}
 -\frac{1}{2}(\gamma_{j}^{\da}+\gamma_{k}^{\da})\sigma_{jk} 
+  \frac{1}{2}(\gamma_{j}^{\ua} + \gamma_{k}^{\ua})\sigma_{jk}
\nextstep
&= \gamma_{k}^{\ua}\delta_{jk}
 -\frac{1}{2}
 \qty( (\gamma_{j}^{\da}-\gamma_{j}^{\ua})
 + 
 (\gamma_{k}^{\da}-\gamma_{k}^{\ua})
 ) \sigma_{jk} 
\nextstep
&=
\gamma_{k}^{\ua}\delta_{jk} - \frac{1}{2}(\gamma_{j}+\gamma_{k})\sigma_{jk} ,
\label{eq_DissipatorPart}
\end{align}
where $\gamma_j = \gamma_j^\da - \gamma_j^\ua$ can be loosely referred to as the relaxation rate~\cite{Alicki_QuantumEngine2023}.
Summing up everything, the dynamics of all of the MDM is governed by:
\begin{align}
\dv{t} \sigma_{jk} =
\frac{i}{\hbar}
\bigg[ \frac{i\hbar}{2} (\gamma_{j}+\gamma_{k}) 
+ \Delta\epsilon_{jk}
+ \sum_m \qty( c_{jm}  \Delta n_{jm} +  c_{mk} \Delta n_{mk} )
\bigg] \sigma_{jk} 
+ \gamma_{k}^{\ua}\delta_{jk} 
 -\frac{i}{\hbar} \sum_{ m } \qty( t_{mk} \sigma_{jm}  - t_{jm} \sigma_{mk} ) .
\end{align}
For later convenience, the evolutions of the populations and coherences can be isolated:
\begin{align}
\dv{t} n_{j} 
=& - \gamma_{j} n_{j} + \gamma_{j}^{\ua} 
+ \frac{2}{\hbar} \sum_{ m \neq j } \Im( t_{mj} z_{jm} ) ;
\nextline
\dv{t} z_{jk} 
=& \frac{i}{\hbar}
\bigg[
\frac{i\hbar}{2} (\gamma_{j}+\gamma_{k}) 
+ \Delta\epsilon_{jk}
+ 2 \sum_m \qty( c_{jm}  \Delta n_{jm} +  c_{mk} \Delta n_{mk} )
\bigg]  z_{jk}
 -\frac{i}{\hbar}  t_{jk} \Delta n_{jk}
 -\frac{i}{\hbar} \sum_{ m \neq j,k} \qty( t_{mk} z_{jm}  - t_{jm} z_{mk} ) 
.
\nonumber
\end{align}
The transient behavior of a device with any desired initial state is revealed through an integration of above set of differential equations, and the steady state is arrived when all the derivatives reach zero.
Alternatively, for directly revealing the steady state one could manually put the derivatives to zero and solve the resultant set of nonlinearly coupled equations [see \Cref{eq:PopCoh_Generic} in the main text].

\twocolumngrid
\bibliographystyle{apsrev4-2}
\bibliography{main}

\end{document}